\documentclass[12pt,preprint]{aastex}
\usepackage{apjfonts}
\usepackage{graphicx}
\shorttitle{On the Mass of CoRoT-7b}
\shortauthors{Hatzes et al. }
\begin{document}
\bibliographystyle{apj}

\title{On the Mass of CoRoT-7b}

\author {Artie P. Hatzes }
\email{artie@tls-tautenburg.de}
\affil{Th\"uringer Landessternwarte, D - 07778 Tautenburg, Germany}
\author{Malcolm Fridlund} 
\affil{European Space Agency, ESTEC, SRE-SA, P.O. Box 299, NL-2200AG, Noordwijk, The Netherlands}
\email{malcolm.fridlund@esa.int}

\author{Gil Nachmani and Tsevi Mazeh }
\affil{School of Physics and Astronomy, Raymond and Beverly Sackler Faculty of Exact Sciences, Tel Aviv University, Tel Aviv, Israel}

\author{Diana Valencia}
\affil{Observatoire de la C\^ote d'Azur, BP 4229, 06304 Nice Cedex 4, France}

\author{Guillaume H\'ebrard}
\affil{Institut d'Astrophysique de Paris, UMR 7095 CNRS, Universit\'e
Pierre \& Marie Curie, 98bis boulevard Arago, 75014 Paris, France}

\author{Ludmila Carone and Martin P\"atzold}
\affil{
Rheinisches Institut f\"ur Umweltforschung, Universit\"at zu K\"oln, Abt. Planetenforschung, Aachener Str. 209, 50931 K\"oln, Germany
}

\author{Stephane Udry}
\affil{Observatoire de l'Universit\'e de Gen\`eve, 51 chemin des Maillettes, 1290 Sauverny, Switzerland}

\author{Francois Bouchy}
\affil{Observatoire de Haute Provence, 04670 Saint Michel l'Observatoire, France}
\author{Magali Deleuil, Claire Moutou, and Pierre Barge}
\affil{Laboratoire d'Astrophysique de Marseille, CNRS \& University of Provence, 38 rue Fr\'ed\'eric Joliot-Curie, 13388 Marseille cedex 13, France}

\author{Pascal Bord\'e}
\affil{Institut d'Astrophysique de Paris, UMR7095 CNRS, Universit\'e Pierre \& Marie Curie, 98bis boulevard Arago, 75014 Paris, France}

\author{Hans Deeg and Brandon Tingley}
\affil{Instituto de Astrof{\'i}sica de Canarias, E-38205 La Laguna, Tenerife, Spain}

\author{Rudolf Dvorak}
\affil{University of Vienna, Institute of Astronomy, T\"urkenschanzstr. 17,
A-1180, Vienna, Austria}

\author{Davide Gandolfi}
\affil{European Space Agency, ESTEC, SRE-SA, P.O. Box 299, NL-2200AG, Noordwijk, The Netherlands}

\author{Sylvio Ferraz-Mello}
\affil{IAG, University of S\~ao Paulo, Brasil}

\author{G\"unther Wuchterl and Eike Guenther}
\affil{Th\"uringer Landessternwarte, D - 07778 Tautenburg, Germany}

\author{Tristan Guillot}
\affil{Universit\'e de Nice-Sophia Antipolis, CNRS UMR 6202, Observatoire de la C\^ote d'Azur, BP 4229, 06304 Nice Cedex 4, France}

\author{Heike Rauer\altaffilmark{1}, Anders Erikson, Juan Cabrera, Szilard Csizmadia}
\affil{Institute of Planetary Research, German Aerospace Center, Rutherfordstrasse 2, 12489 Berlin, Germany}

\author{Alain L\'eger}
\affil{Institut d'Astrophysique Spatiale, Universit\'e Paris XI, F-91405 Orsay, France}

\author{Helmut Lammer and J\"org Weingrill}
\affil{Space Research Institute, Austrian Academy of Science, Schmiedlstr. 6
A-8042, Graz, Austria}

\author{Didier Queloz and Roi Alonso}
\affil{Observatoire de l'Universit\'e de Gen\`eve, 51 chemin des Maillettes, 1290 Sauverny, Switzerland}
\author{Daniel Rouan}
\affil{LESIA, Observatoire de Paris, Place Jules Janssen, 92195 Meudon cedex, France}

\author{Jean Schneider}
\affil{LUTH, Observatoire de Paris, CNRS, Universit\'e Paris Diderot; 5 place Jules Janssen, 92195 Meudon, France}

\altaffiltext{1}{Center for Astronomy and Astrophysics, TU Berlin, Hardenbergstr. 36, 10623 Berlin, Germany}

\begin{abstract}
The mass of CoRoT-7b, the first transiting superearth exoplanet, is still a subject of debate. A 
wide range of masses have been reported in the literature ranging from as high as 8 $M_\oplus$
to as low as 2.3 $M_\oplus$. Although most mass
determinations give a 
density consistent with a rocky planet, the lower value
permits 
a bulk composition that can be up to 50\% water. 
We present an analysis of the CoRoT-7b radial
velocity measurements that uses very few and 
simple assumptions in treating the activity
signal.  By only analyzing those radial velocity data for which multiple measurements 
were made in a given night we remove the activity related
radial velocity contribution without any a priori
model. We demonstrate that the contribution of  activity to the final radial
velocity curve is negligible and that the $K$-amplitude due to the planet is well constrained.
This yields a mass of 7.42 $\pm$ 1.21 $M_\oplus$ and a mean density of 
$\rho$ = 10.4 $\pm$ 1.8 
gm cm$^{-3}$. 
CoRoT-7b is similar in mass and radius to the second rocky planet to be
discovered, Kepler-10b, and within the errors they have identical bulk densities - they are virtual twins. These
bulk densities lie close to  the density - radius relationship for terrestrial planets similar to what is seen for Mercury.
CoRoT-7b and Kepler-10b may have an internal  structure more like Mercury than the Earth. 
\end{abstract}


\section{Introduction}
\label{intro}

The discovery of the first superearth with a {\it measured} absolute radius and mass was recently reported (L\'eger et al.~2009; Queloz et al.~ 2009; 
Hatzes et al.~2010). This detection was based on the photometric measurements made by the CoRoT satellite (Convection, Rotation and planetary Transits, Baglin et al.~2006). What made this detection so interesting was that the average densities calculated from the very accurate radii and radial velocity measurements of these authors all indicated values  between 5.7 $\pm$ 1.3 and 9.7 $\pm$ 2.7 gm cm$^{-3}$. When taking the actual radius into account, these values are indicative of similar values found for the terrestrial planets in the Solar System (Mercury, Venus and the Earth).

In the exoplanet community there has been some discussion regarding the nature of 
CoRoT-7b. Is this a rocky planet with a density consistent with terrestrial planets 
(Queloz et al.~2009; Hatzes et al.~2010), or
is it a volatile-rich planet? 
The reason for this uncertainty
in the possible composition  is due to the wide range of planet masses that have been reported in the
literature. Queloz et al.~(2009) give a mass of CoRoT-7b of $m$ = 4.8 $\pm$ 0.8 $M_\oplus$, 
Hatzes et al. 2010 report a mass of $m$ = 6.9 $\pm$ 1.5 $M_\oplus$, 
Ferraz-Mello et al. (2010) a mass of
$m$ = 8.0 $\pm$ 1.2 $M_\oplus$, and Boisse et al. (2011) a mass
of $m$ = 5.7 $\pm$ 2.5 $M_\oplus$. 
While most authors favor a relatively high value for the planetary mass (and thus density) that is consistent with 
a rocky composition,  
on the low mass end Pont et al (2010) give a mass of $m$ = 2.3 $\pm$ 1.8 $M_\oplus$ and with a rather low 
significance of detection of the planet of 
only 1.2$\sigma$. This low value excludes a rocky composition and can only be reconciled with relatively 
large amounts of volatiles ( $\sim$30\% of water by mass in vapor form, Valencia 2011).
Pont et al. (2010) claim that given the activity
signal of the host star, the actual mass of CoRoT-7b 
can range from 1--5 M$_\oplus$ and thus allow a large range of bulk compositions.  
They caution the reader about building models based on the 
rocky nature of CoRoT-7b.
The perceived ``uncertainty'' of the mass of CoRoT-7b seems to linger in spite of the
excellent quality of the radial velocity (RV) measurements taken with the High 
Accuracy Radial velocity Planet Searcher (HARPS) spectrograph (Mayor et al.~2003).
Removing this uncertainty is
essential if theoreticians are to construct valid models of the internal structure of
CoRoT-7b.

The reason for the wide range of mass estimates for CoRoT-7b stems from
the fact that the host star, CoRoT-7 is  relatively active. The CoRoT light curve shows 
a modulation of $\approx$ 2\% with a rotation period  of 23 days. This light
curve also shows clear evidence of spot evolution over the 132 day observing
window of CoRoT. The RV ``jitter'' due to 
activity is much larger
than the expected planet signal. Further complicating the analysis 
is the
presence of a second (Queloz et al.~2009) or possibly even a third planetary companion
(Hatzes et al. 2010) with periods of 3.7 and 9 days, respectively. 
The planet RV amplitude, central to
determining the companion mass, depends on the details of 
how the activity signal 
is removed. We note that all the above mass determinations utilized the same
HARPS RV data set which only emphasizes the challenge in determining the planet mass for an active star like CoRoT-7. 
However, we should point out that 
CoRoT-7 is not {\it that} active a star, particularly when compared to T Tauri, RS Cvn-type, and very young
stars - classes of objects considered to be very active. 
In terms of mass, radius, effective
temperature, rotational period,  and photometric variations it is more
similar to the Sun than say the planet hosting star  CoRoT-2 (Alonso et al.~2008).
We therefore can use  solar activity as a good proxy for understanding the behavior of
activity in CoRoT-7b.

The HARPS data used for all of these mass determinations
 consisted of a total of 106 precise RV measurements acquired over four 
months with a typical error of $\approx$ 2 m\,s$^{-1}$ (Queloz et al.~2009).
The spectral data 
also produced useful activity indicators that included Ca II  emission,
the bisector of the cross-correlation function (CCF), and the full width
at half maximum (FWHM) of the CCF.

A variety of techniques have been employed to extract the planet RV signal from that due
to activity. Queloz et al.~(2009) presented two approaches. The first method
used harmonic filtering of the data using the  CoRoT photometric 
rotation period, $P_{rot}$  of 23 days and its three harmonics of
$P_{rot}$/2,   $P_{rot}$/3, and $P_{rot}$/4. This resulted in 
an amplitude of 1.91 $\pm$ 0.04  m\,s$^{-1}$. A Fourier analysis using pre-whitening 
(cleaning) resulted in a higher amplitude
of 4.16 $\pm$ 1.0  m\,s$^{-1}$.  (Although the internal error was $\pm$ 0.27 m\,s$^{-1}$,
Queloz et al.~2009) argued that because of the pre-whitening process a more realistic error
was $\pm$ 1.0 m\,s$^{-1}$ which we use here.)
After correcting for possible effects of the various filtering processes a common amplitude of
3.5 $\pm$ 0.6  m\,s$^{-1}$ was adopted.  Ferraz-Mello et al. (2010) used 
a self consistent version of harmonic filtering (denoted ``high pass filtering'') to get a
$K$-amplitude of 5.7  $\pm$ 0.8 m\,s$^{-1}$. Boisse et al. (2011) also used a version of harmonic filtering
of the HARPS data to derive a $K$-amplitude of 4.0 $\pm$ 1.6  m\,s$^{-1}$ .
Hatzes et al. (2010) used orbital fitting using only data from 
nights where more than one RV measurement was made. This resulted in a $K$-amplitude of 5.04 $\pm$ 1.09 m\,s$^{-1}$.

 In this paper we present a simple approach to removing the activity signal that is model
independent. We use  
a subset of the HARPS data that has
multiple measurements per night and allow the nightly offsets to vary
(Hatzes et al. 2010). This results in a $K$-amplitude  that is not affected largely by
the activity signal and that is consistent with other recent determinations.
This simple 
``filtering''
does not assume a  rotation period for the star, nor  does it require the use of any of its harmonics.
This procedure confirms the high mass of CoRoT-7b and that bulk compositions consisting of up to 50\% water can be reliably
excluded.  We compare the density of CoRoT-7b 
to that of the recently reported Kepler-10b (Batalha et al. 2011).
What is so striking about these two planets is they orbit
very similar stars (G9V and G4V respectively), at comparable orbital  distances 
(both planets have orbital periods of $\approx$ 20 hours), and they have a similar radius and
mass. The only difference is in stellar metalicity, as well as in the activity level.
 CoRoT-7 shows very high levels of activity, 
while Kepler-10b is one of the quietest stars in the Kepler sample. 
Furthermore, the planet Kepler-9d 
(Holman et al. 2010; Torres et al. 2011), has  a similar well-determined radius ($\approx$ 1.6 R$_{\oplus}$), a slightly longer period (1.59 d) and an upper mass limit of 7 M$_{\oplus}$, which if confirmed would bring it into the same density regime.

\section{Removing the activity RV signal}

Hatzes et al. (2010) used a simple method for removing the RV signal due to activity. 
The method exploits the fact that the RV variations due to the transiting
planet have much shorter timescales than those expected from 
the activity signal. Only  those HARPS data 
that had multiple measurements taken on the same night and with good time 
separation between successive measurements were used. This resulted in a 
total of 64 measurements or about one-half of the HARPS RV
data (a case of less being more). (Note that this includes an additional
4 nights of data inadvertently left out by Hatzes et al.~(2010)). Table 1
lists the starting Julian Day for the nightly measurements, the number of measurements taken on that night, 
the time separation between first and last nightly measurement, and the resulting orbital phase difference between the first and last observation on that
night.  
An orbital solution was made using the generalized non-linear least squares program 
{\it Gaussfit} (Jeffereys et al.~1988)
and treating data from individual nights as independent data sets. The period was fixed to the CoRoT-determined transit value of 0.85-days, but the phase, amplitude, and nightly offsets
were allowed to vary.
Initially, the eccentricity was fixed  to zero.
There are two basic and very simple assumptions
in this analysis (hereafter refereed to as ``model independent''): 
1) A 0.85 day RV period is present in the data. 2)
The RV zero point offset does not vary significantly
 over the span of measurements for a single night.

Assumption 1) is reasonable given the clear transit signal in the data. L\'eger, Rouan, Schneider et al.~(2009) carefully excluded
all possible false positives and established with a  high degree 
of confidence that a planetary companion was causing the transit event. Hatzes et al. (2010)
showed that even this assumption can be relaxed as the 0.85-d period 
provides the best fit to the data.  Furthermore, multiple RV measurements taken on a
given night show short term variability consistent with the presence of a short period.

Assumption 2) is also eminently reasonable. The rotation period is 
known to be 23 days and over the time span of the nightly observations 
(maximum $\approx$ 4 hours) 
the star rotates by  no more than 2.6 degrees. The RV amplitude due to 
activity rotational modulation 
is $\approx$ 10 m\,s$^{-1}$ (Queloz et al.~2009). This
means that the change in RV due to rotational modulation  will amount
to a {\it maximum} value of $<$ 0.5 m\,s$^{-1}$.
Rapid and significant changes 
in the spot distribution on time scales of $\approx$ 4 hours in a star
with solar-like activity like CoRoT-7 would 
be unprecedented. For instance, measurements of the solar constant
from the Solar Maximum Mission shows peak variations of 0.2\% on
timescales of {\it weeks} (Foukal~1987). In short, the RV contribution due to activity over the time
span of the nightly measurements should be nearly constant. 

If there are additional planets present then these might contribute to
changes in the mean RV for a given night that would not subtract
out fully. 
These, however, also make a negligible  contribution. For the sake of argument let us 
assume that CoRoT-7c and d are also present. Using the orbital solutions
of Hatzes et al. (2010) we can calculate the change in RV on a given night
due to the reflex motion of the star caused by these companions.  This
is listed in the fourth column of Table 1 under $\Delta T_{C7c,d}$. (Note that due
to the shorter period the largest effect stems from CoRoT-7c.) The average
change in RV is $-$0.04 $\pm$ 0.89 m\,s$^{-1}$. The maximum change in RV of
2 m\,s$^{-1}$ only 
adds a maximum possible  error of $\pm$ 1 m\,s$^{-1}$ to the zero-point offset. For most nights
the error is less than 0.5--1 m\,s$^{-1}$.
Combined with the rotational modulation signal, the error in the nightly
RV zero point offsets (mean values) is less than the measurement error of about
2 m\,s$^{-1}$. 

We stress that in this procedure we essentially do not care what the contribution
of surface spots or other planets are to the RV signal. By 
calculating the mean RV value for a given night (in a least
squares sense) and subtracting this from the signal we effectively remove
all other contributions to the RV signal that have periods substantially longer than
the orbital period of CoRoT-7b.
Our approach will also
remove any unknown long-term systematic instrumental errors. 
The stability of the 
HARPS spectrograph is nearly legendary with nightly drifts of 
the spectrograph on average being less than 0.5 m\,s$^{-1}$ (Lo Curto et al. 2010). Even if HARPS had night-to-night 
systematic errors, our approach should eliminate, or at least
greatly minimize these. As  long as the RV variations from other sources do not
change significantly over a 2--4 hour time span, any RV variations that are seen
in a given night can be attributed solely to that of CoRoT-7b.

Figure~\ref{corot7} shows the nightly RV measurements after removal of the nightly zero-point 
offsets and phased to the CoRoT
transit ephemeris.  Initially, the eccentricity was fixed at zero as was the
ephemeris and orbital period. 
Allowing the phase to vary resulted in a value to within a phase
of 0.01 to the CoRoT value. Varying 
the orbital eccentricity  resulted in a
small eccentricity ($e$ = 0.077) but with large errors ($\pm$ 0.11). The orbit is circular to within the error. The solution shown is for zero eccentricity and phase
fixed to the transit phase. The resulting $K$-amplitude is 
5.15 $\pm$ 0.95 m\,s$^{-1}$

When examining the data taken in a single night 
one clearly sees that the change in the RV almost always follows the orbit 
to within the measurement error. The rms scatter
of the RV data about the orbital solution (line) is 1.68 m\,s$^{-1}$ which is in excellent
agreement with the median RV internal error of 1.77 m\,s$^{-1}$. This figure alone argues that the contribution of 
activity jitter to this RV curve is negligible. 

As a ``sanity check'' we performed a Scargle periodogram (Scargle~1982) on the  calculated
nightly offsets. This periodogram, shown in Figure~\ref{offset}, has its highest
peak at a  frequency of $\nu$ = 0.043 c\,d$^{-1}$, 
the known rotational frequency of the star. 
According to Scargle~(1982) the probability that noise will produce
this peak exactly at the rotational frequency is about 
5$\times$10$^{-4}$. This was confirmed using a bootstrap randomization 
procedure.
The RV data was randomly shuffled 200,000 times,  keeping the times fixed and the maximum power in the random data periodogram was examined.
Over the narrow frequency range 0.035--0.055 c\,d$^{-1}$ centered on the
rotational frequency 
the false alarm probability that noise is causing the observed peak was 7$\times$10$^{-5}$.
The derived nightly 
offsets thus have some relationship to a known phenomenon associated with the 
star (i.e. rotation). The RV amplitude of this peak is $\approx$ 10 m\,s$^{-1}$, consistent
with the RV amplitude due to rotational modulation using the entire HARPS
data set (Hatzes et al. 2010).

We performed Monte Carlo simulations to confirm the error on the $K$-amplitude,
but more importantly to assess how well we could recover a known signal in the
RV data. Given the relative sparse sampling of the sine-curve on each night there
was some concern that when fitting a fixed period to the data the procedure may
introduce variations with that period into the data due primarily to the freedom in
varying the nightly offsets. 
A synthetic sine wave was generated with an amplitude of 5 m\,s$^{-1}$
and having the same period and phase of CoRoT-7b. The fake data was sampled in the same manner as the
real data and random noise at a level of 2 m\,$^{-1}$ was added. To mimic the
activity signal random noise with an rms of 10 m\,s$^{-1}$ was  added to the fake
RVs for each night.
 The data was then processed in the same manner as the real data. 
One hundred such simulations yielded a mean amplitude of 5.00 $\pm$ 0.94 m\,s$^{-1}$ 
and in all
cases the nightly ``activity'' signal was recovered to within the measurement
error. The error in the $K$-amplitude from the simulations is entirely consistent
with the measured value and suggests that the errors in the RV measurements
are nearly Gaussian.

Figure~\ref{orbit} shows the RV measurements after they have been
 averaged in bins
of approximately $\Delta\phi$ $\sim$ 0.05--0.1. The nightly
zero-point offsets calculated using the unbinned data have been applied.
We have plotted the abscissa and ordinate on the same scales used 
for the binned RV curve for Kepler-10b shown in Figure 6 of Batalha et al.~(2010) so as
to facilitate a direct comparison between the RV curves of these two superearths.
This  figure shows that in spite of an RV ``jitter'' of  10 m\,s$^{-1}$ for
CoRoT-7, that with better temporal sampling and our simple treatment of the
activity signal we can extract an RV curve that is much ``cleaner'' than for a
quiet star such as Kepler-10b.  We do have more free parameters in our fit - the nightly 
offsets due to activity - but by taking several measurements per night we can determine
and correct for the RV jitter. This, of course, would be impossible with just a single RV measurement per
night (or for a planet with a much longer orbital period). 
Calculating an orbit using the binned values and only 2 free parameters
- the $K$-amplitude and a single
zero-point velocity for all nights - results in an amplitude  of
$K$ =  5.21 $\pm$ 0.21 m\,s$^{-1}$. We believe this 
error to be artificially low because we are no longer fitting different nightly zero-points
as we did with the unbinned data and these errors are correlated with the 
error in the $K$-amplitude. So, the error of $\sigma$ = $\pm$ 0.94 m\,s$^{-1}$ 
including the correlated errors of the nightly offsets is a more realistic assessment
of the error in the $K$-amplitude.
The rms scatter of the RV measurements
about this orbital solution is 0.52 m\,s$^{-1}$.

\section{Linear Model}

Our Keplerian solution from the previous section established that the orbit of
CoRoT-7b is circular and in
phase with transit lightcurve. We can thus use 
a {\it linear} model of the form:
\begin{equation}
Acos(\omega t_i) + Bsin(\omega t_i) + C_i,
\end{equation} 
where A and B determine the amplitude and phase of the sine function fitted to the RV periodic modulation, and  $C_i \in \left \{  \right. C_1,...,C_{27}\left.\right\}$ are the 
different shifts for each of the 27 nights with multiple measurements (each $C_i$ is assumed to be constant for each night), $\omega =2\pi/P $ for a given period $P$ and $t_i$ is the time of each measurement. 
Presenting the model in the form of Equation (1) enabled us to apply a linear model to the data as the
phase does not appear explicitly. The linear model assured a robustness 
of the best solution while
The resulting radial velocity amplitude from the linear model was 5.25  $\pm$ 0.84 m\,s$^{-1}$.
We shall use this value as our ``best-fit'' $K$-amplitude.

We also used the linear model as a ``high pass filter'' and applied 
it to the data using different frequencies.
The periodogram of $1/\chi^2$ calculated with fits using  frequencies in the 
range (0,2] days$^{-1}$ (Figure~\ref{periodogram}) shows a clear global maximum near 
Corot-7b's period, 
which was found through photometry by Leger et al.~(2009). 

\section{Derivative Fitting}

The derivative of the RV curve should also
be insensitive to the contribution of the activity and taking differences
in the RV on a given night one should only detect changes due to orbital motion and the
underlying (constant) activity signal removed.
The  fitting to the RV data done in the previous sections confirms that the orbit is circular
and in phase with the CoRoT ephemeris. The RV curve can thus be described by the
function: $V$  = $-Ksin(2\pi\phi)$ $+$ $C_i$, where $K$ is the RV amplitude, $\phi$ the
orbital phase, $C_i$ is the nightly zero point, and the minus sign ensures that
 the RV curve has the proper phase with respect to the mid-transit time.
Differentiating
with respect to $\phi$ results in $dV$/d$\phi$ = $-2\pi K$cos($2\pi\phi$).
So, a plot of the derivative (differences) of the RV values versus
cos($\phi$) should have a linear relationship whose slope is proportional
to the $K$-amplitude.

Figure~\ref{deriv} shows the normalized nightly RV derivatives ($dV/d\phi$/2$\pi$)
as a function of $-$cos($\phi$). There is considerable scatter since we are
dealing with differences, but there is a clear correlation.
The correlation coefficient is $r$ = 0.59
with a probability of 6 $\times$ 10$^{-5}$ that the data is uncorrelated.
A least squares fit to the derivatives yields a $K$-amplitude of
5.02 $\pm$ 1.25 m\,s$^{-1}$ a value entirely consistent with the previous
approaches.

\section{Contribution of the Activity Signal}

Table 3 lists all the different
$K$-amplitude determinations for  CoRoT-7b from here and the literature.  
For Queloz et al.~(2009) we
include not only the adopted value, but the results from the individual methods
for determining the RV amplitude. To ensure that the mass
was determined in a consistent way using the same stellar and orbital parameters we recalculated
these using the published $K$-amplitudes and the mass function
(see below). These may result in slightly different
values than quoted in the respective papers. These appear in the third column in the table
and graphically
in Figure~\ref{masses}. The number on the abscissa corresponds to that in the first
column of the table. 
Our mass for CoRoT-7b  agrees  to within 1$\sigma$ to most other mass determinations.
It is  most discrepant (4$\sigma$) with the harmonic filtering without correction
from Queloz et al.~(2009), but the authors showed that the filtering process 
resulted in an unusually low value. The authors considered the adopted value of
4.96 $\pm$ 0.86 to be a more accurate determination.

Our mass value is clearly 
discrepant with the low value of 2.3  $\pm$ 1.8 M$_\oplus$ (2.1$\sigma$)
claimed by Pont et al. (2010), a mass determination largely responsible 
for casting
doubt on the rocky nature of CoRoT-7b (Batalha et al. 2011).
We believe the 
Pont et al. value to be
artificially low, most likely due to their treatment of the activity.
In modeling the RV and activity indicators they used 2-200 spots
($N$=12-20 could reproduce the light curves within the uncertainties). These
spots were parameterized by a scale factor (no spot temperature was specified), 
a latitude, and a longitude. The
spot evolution was characterized via a Gaussian  with parameters of peak
intensity, epoch, and lifetime. Differential was included, although values
were not specified. A relatively large number of parameters and assumptions were
used for fitting the observed quantities. 

\begin{figure}
\plotone{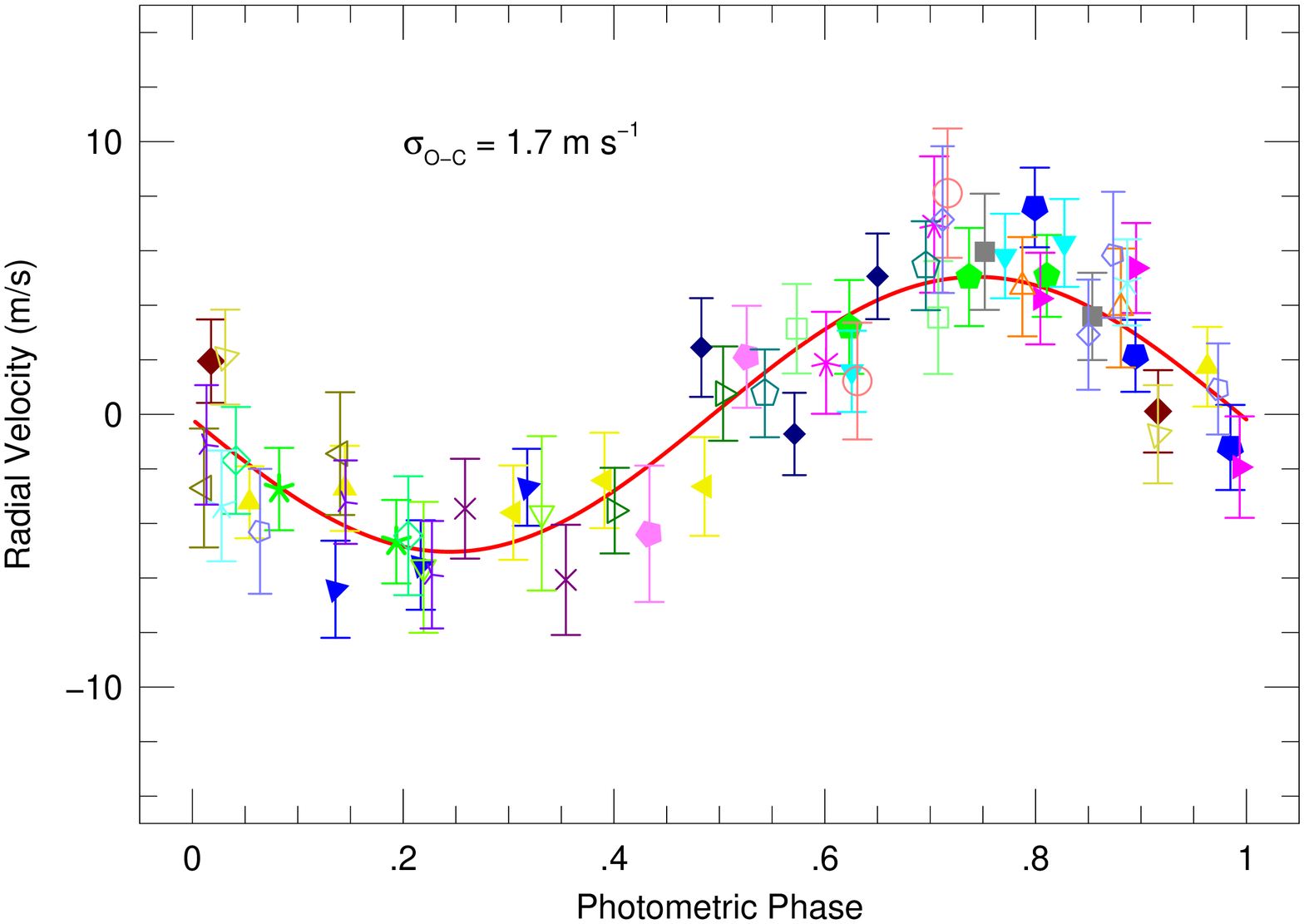}
\caption[]{RV measurements for CoRoT-7b taken on the 27 nights listed
in Table 1. On these nights multiple observations were made separated by
2--4 hours. Values are phased according the CoRoT transit phase. Different
symbols represent measurements for different nights. The line represents
the orbital solution. 
\label{corot7}}
\end{figure}

\begin{figure}
\plotone{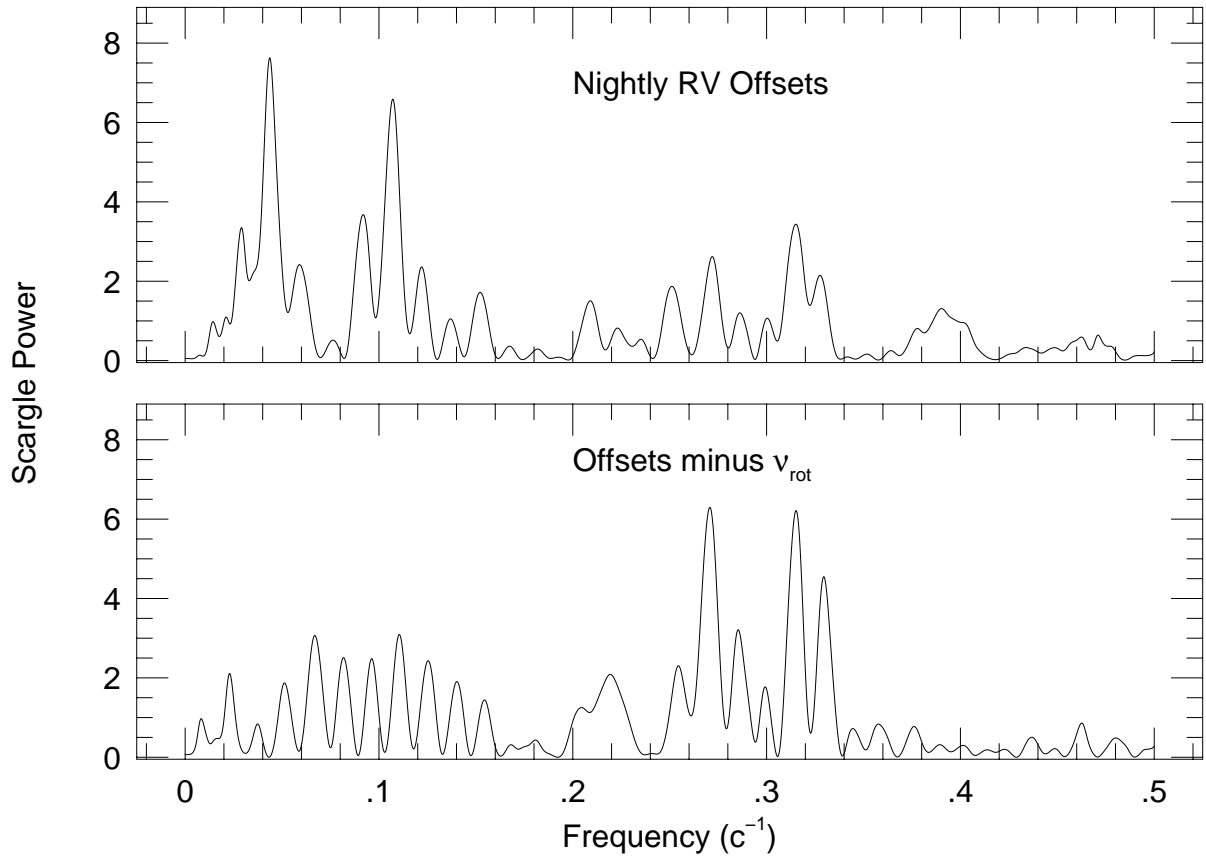}
\caption[]{(top) The Scargle periodogram of the nightly offsets that were determined
in the orbit fitting. The peak corresponds to the stellar rotational frequency of 0.043 c\,d$^{-1}$.
(bottom) The Scargle periodogram of the offset residuals after removing the contribution of the
dominant peak.
\label{offset}}
\end{figure}

\begin{figure}
\plotone{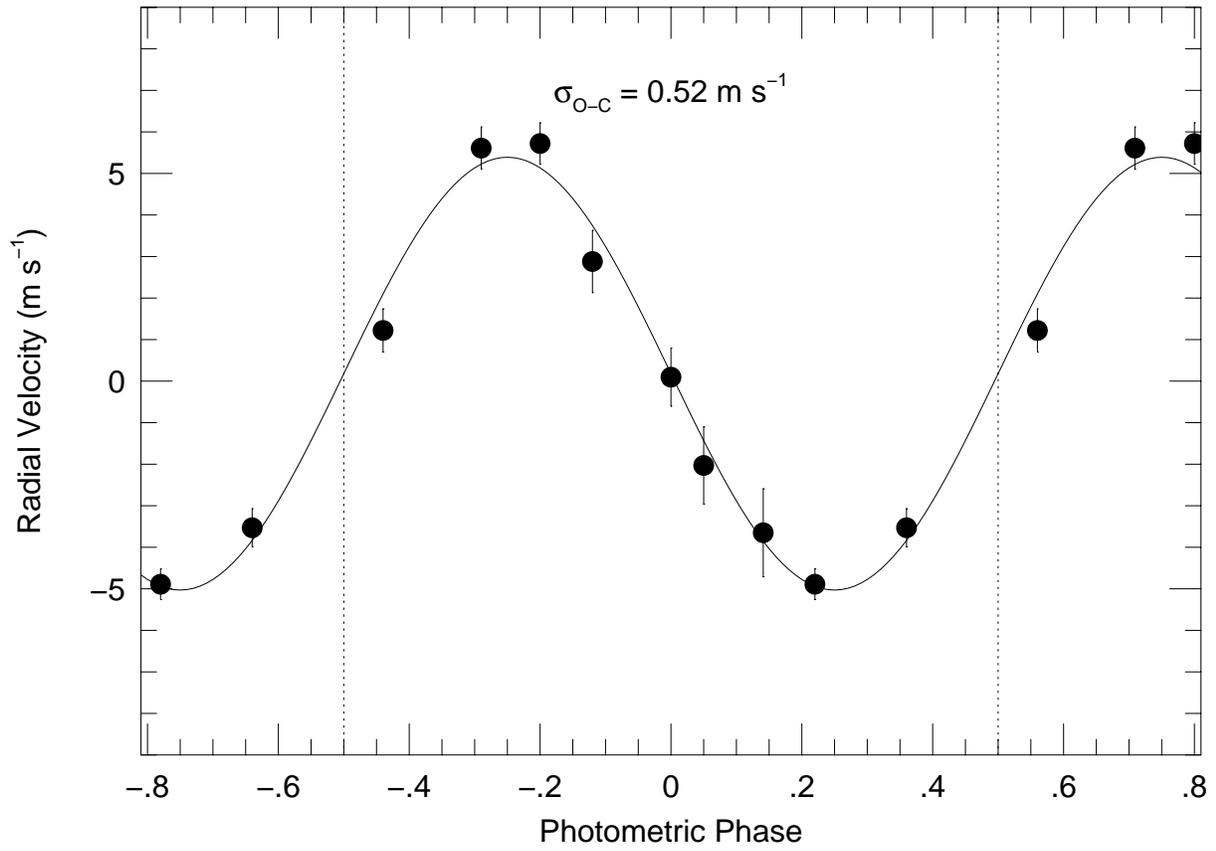}
\caption[]{The phased RV measurements binned by units of $\Delta\phi$ 
$\approx$ 0.1. The line represents the same orbital solution as Figure 1.
The rms scatter
of the data values about the fit is 0.52 m\,s$^{-1}$
\label{orbit}}
\end{figure}

\begin{figure}
\plotone{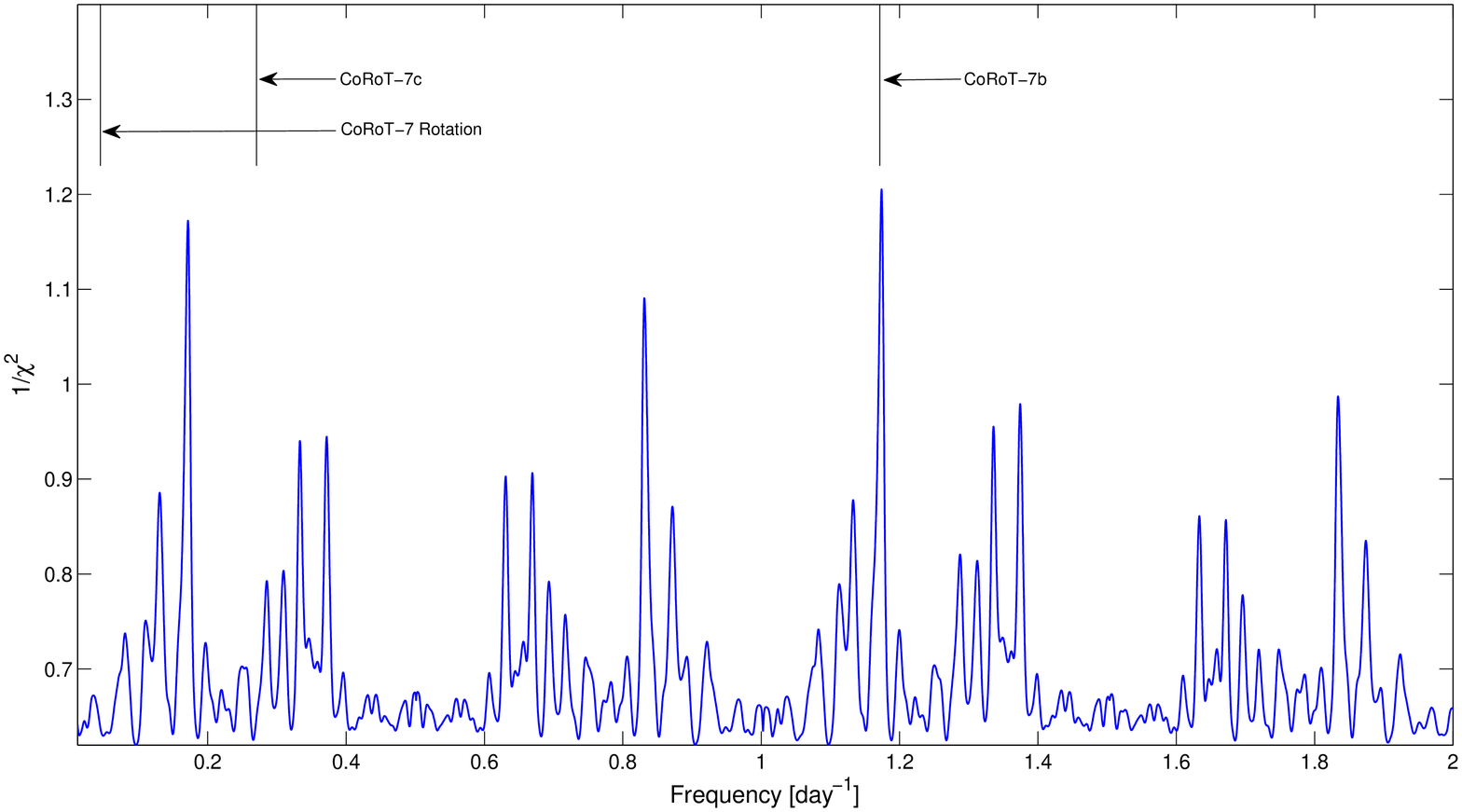}
\caption[]{Periodogram of 1/$\chi^2$ versus frequency calculated using the linear model.
\label{periodogram}}
\end{figure}

\begin{figure}
\plotone{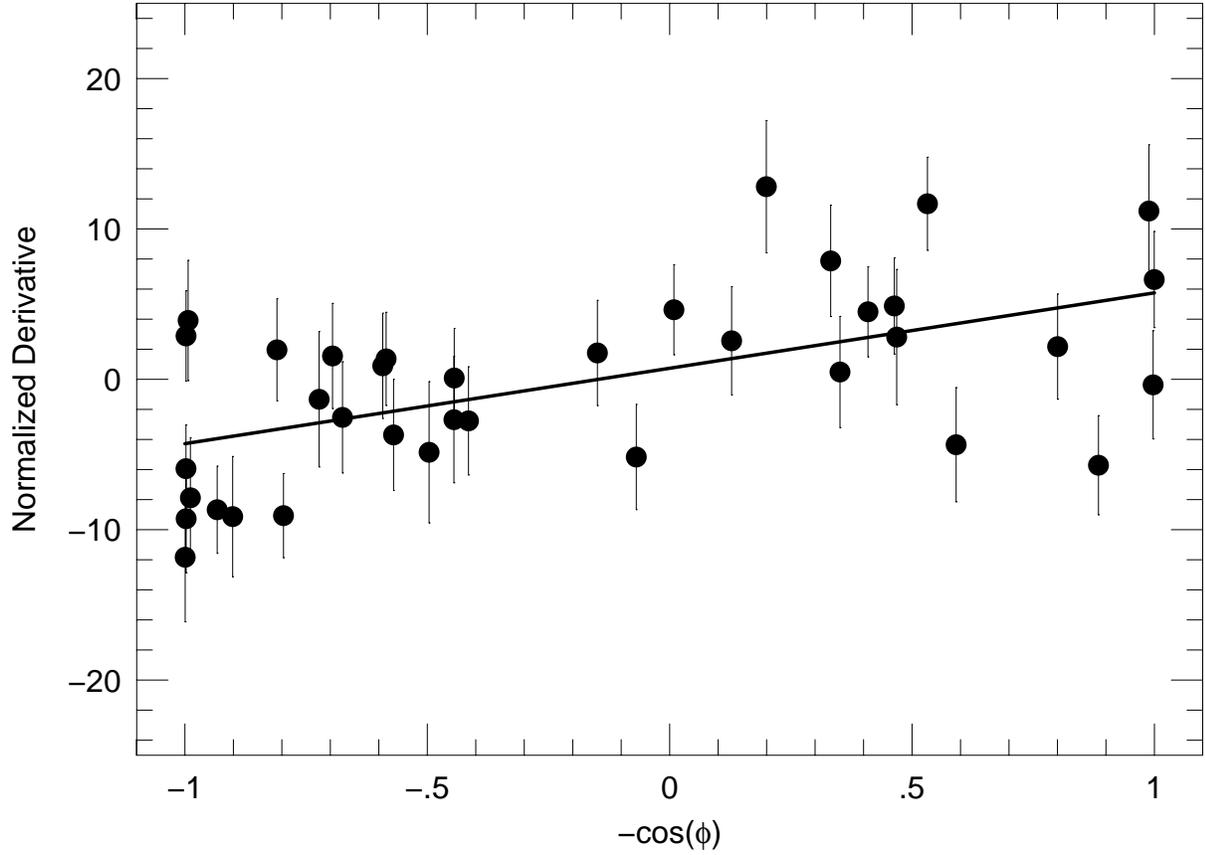}
\caption[]{The normalize derivative, ($dV$/$d\phi$)/$2\pi$) calculated
from successive RV measurements plotted versus
the negative of the cosine of the orbital phase.
The data have a correlation coefficient of $r$ = 0.59 and with the probability
of 6 $\times$ 10$^{-5}$ that they are uncorrelated. The slope  corresponds to 
an RV amplitude of $K$ = 5.02 $\pm$ 1.26
m\,s$^{-1}$.
\label{deriv}}
\end{figure}

\begin{figure}
\plotone{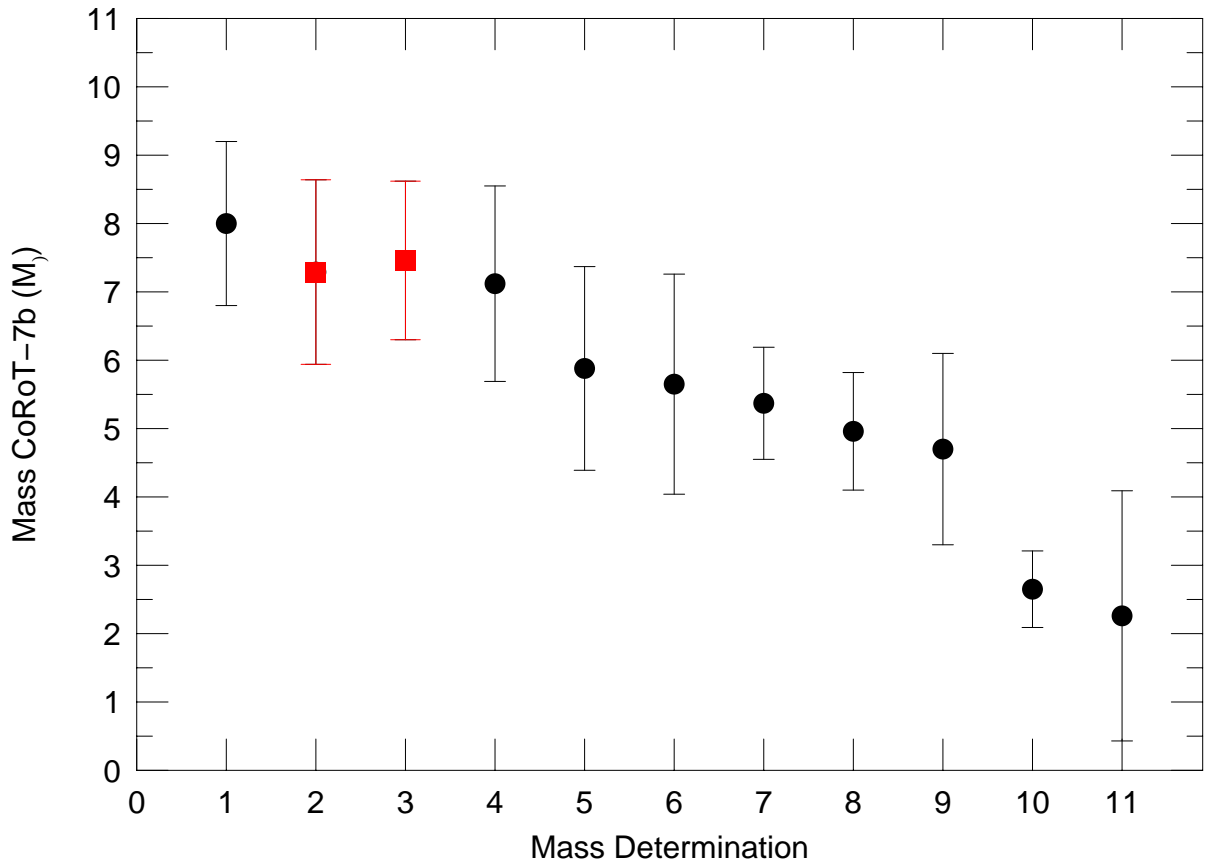}
\caption[]{A comparison of the masses for CoRoT-7b that have been reported in the literature.
The values on the abscissa  correspond to numbers in Table 3. The squares mark the determinations
in this work.
\label{masses}}
\end{figure}

A better understanding of the nature of the 0.85-d RV variations of CoRoT-7b
comes not from some ad hoc activity model, but rather from the RV curve itself.
The striking feature about these curves (Figs.~\ref{corot7}  and \ref{orbit})  is that they
do  not 
deviate from a circular orbit (i.e. pure sine curve) 
to within the errors {\it using the period and phase of the photometric transit}. We have 3 possibilities
for the shape and amplitude of this sine curve: 1) it is due purely to a planet,
2) it is due purely to activity, or 3) it is due to a combination of both.

We can exclude points 2) and 3). For 
activity to contribute significantly
to the RV curve it would have to have variations that would mimic the 
0.85-d period and be {\it perfectly in phase with the CoRoT-7b lightcurve
ephemeris}, otherwise we could not fit a circular orbit to the data. 
We could concoct a spot distribution that evolves with just the right
time scales and phase with respect to the 0.85-d period, but that would 
be too ad hoc and lacking in a reasonable physical interpretation.
The simplest and most logical way to  produce a 0.85-d variation in the RV data,
and one that has a stronger physical basis, 
is with the 1-day alias of $P_{rot}/4$
($\nu$ = $\nu_{rot}$/4 + 1) which is close to the CoRoT-7b orbital frequency.
Figure 8 of Hatzes et al. (2010)  and  the periodogram in Figure~\ref{periodogram}
demonstrate that the
true period of 0.85-d is  favored over the alias period and that it provides
a better fit to the data. 
However, for the sake of
argument let us {\it assume} that the RV curve in Figure~\ref{orbit} is indeed due
to an alias of 4$\nu_{rot}$ rotational harmonic or at least makes
a significant contribution to this curve. We argue that this cannot be the
case for two strong reasons. 

\begin{figure}
\plotone{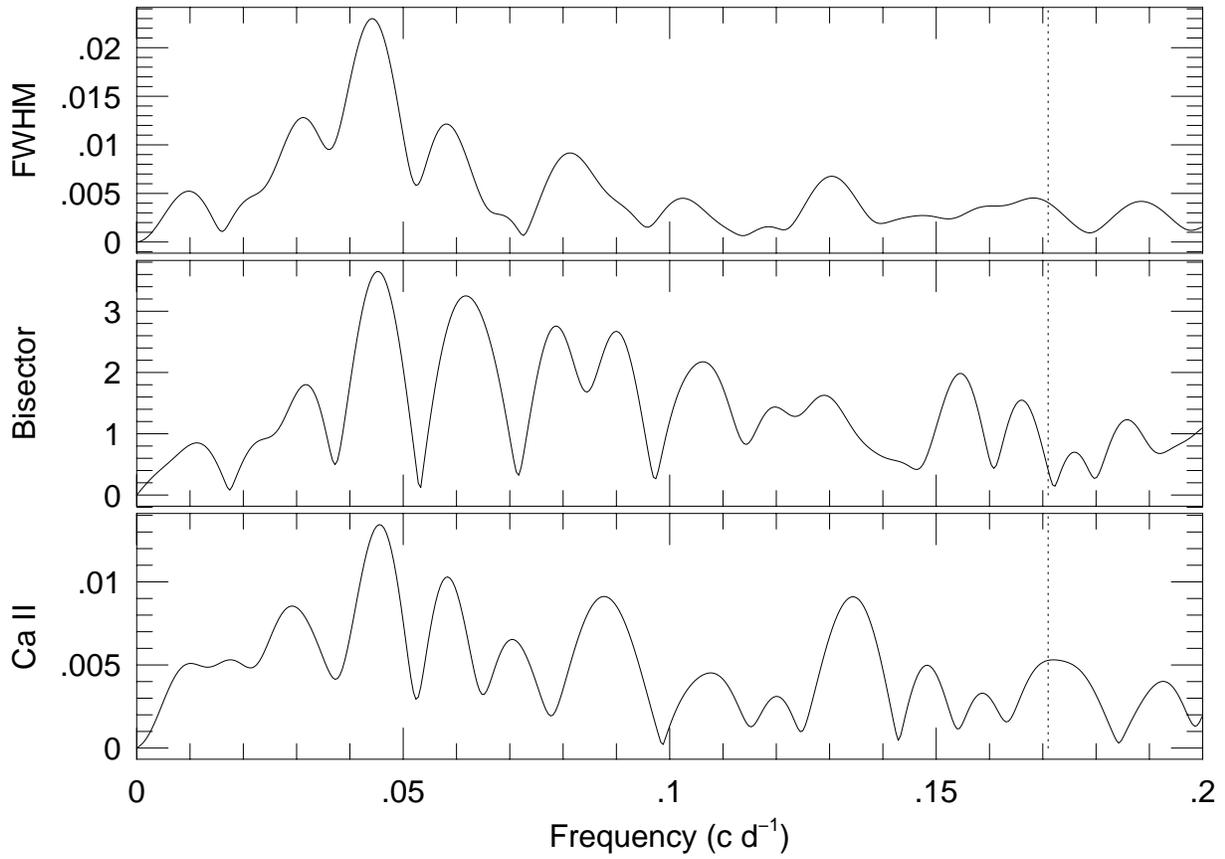}
\caption[]{The amplitude spectrum of the FWHM (top), bisectors (middle),
and Ca II emission (bottom) calculate using the discrete Fourier transform.
The vertical dashed line marks the location of the 4$\nu_{rot}$ 
which is the one-day alias of the
orbital frequency of CoRoT-7b.
\label{indicators}}
\end{figure}

First, there is little evidence for the 4$\nu_{rot}$ actually being present
in the activity indicators. Assuming that the Pont et al. $K$-amplitude of
1.6 m\,s$^{-1}$ is indeed correct, then the RV amplitude due
to activity in Figure~\ref{orbit} amounts to 3.4 m\,s$^{-1}$. The RV amplitude
associated with the rotational frequency is
8.7 m\,s$^{-1}$. This means that the 
third harmonic should have an amplitude $\approx$ 0.4 of the rotational peak.
We find, however, weak evidence for significant power at 4$\nu_{rot}$.
Figure~\ref{indicators} shows the discrete Fourier transform (DFT) amplitude spectra
of the FWHM, the bisectors, and the Ca II emission. The vertical line marks
the the location of $\nu_{rot}$/4 which is near the one-day
alias of the orbital frequency of CoRoT-7b. Only the FWHM and
Ca II show a bit of power near this rotational harmonic. The FWHM shows a ratio of the
third harmonic to rotational frequency of only 0.2. 
Ca II does show a peak at the
third harmonic  with the  proper amplitude ($\approx$ 0.4). However, the DFT of both
the FWHM and bisector span
are  quite noisy with other peaks
close to the third harmonic having a larger amplitude. 
Pont et al. (2010) argue
that the 
FWHM can be used to reconstruct the brightness variations to 0.1\% or better.
If this is true, and since 4$\nu_{rot}$
has an amplitude 20\% of the rotational peak, then 
the RV variations due to the third harmonic 
should be about 1.7  m\,s$^{-1}$, or comparable to the measurement error.

Second, and just as importantly, we would have to place very stringent constraints on the
spot distribution and its evolution  that are highly unlikely:

\begin{enumerate}

\item Physically, the only way to produce a  strong 
presence of $P_{rot}$/4 in the
RV rotational modulation is to have 4 spot groups equally spaced in longitude
by 90$^{\circ}$. Each spot group would produce its own sinusoidal variation
and all four RV curves from the individual spot groups would add together
to contribute to the observed 0.85-d RV variations (in this case,
the one day alias
of $P_{rot}$/4.)

\item For activity to reproduce the CoRoT transit phase and to be  able to add or 
subtract to the RV curve in the proper phase, 
 one spot group {\it must  be
located at transit phase zero}. Otherwise the activity signal would introduce
significant distortions to the RV curve. 

\item To produce the symmetrical RV curve in Figure~\ref{orbit}  and with little 
scatter all four spot groups
must have comparable filling factors, otherwise
the envelope of RV curves from the individual spot groups would introduce significant
``jitter'' and the observed scatter of the RV data would exceed the measurement errors.
For example, this is seen in the RV curve for Kepler-10 where the assumed RV jitter dominates
the internal measurement error in spite of this star being  ``quiet'' in terms
of activity. (Batalha et al. 2011).  Here the resulting binned RV curve is more distorted and
``noisier''
compared to that of CoRoT-7.
This filling factor can be estimated 
using the expression of Hatzes~(2006) or Saar \& Donahue~(1996). 
To produce  a spot-induced RV amplitude
of 5  m\,s$^{-1}$ requires a spot filling factor for each group of about 0.3 \%. The difference
in spot filling factor (areal coverage)
can be estimated from the rms scatter about the orbital solution.
For the RV curves from the different spot groups to produce variations 
less than the rms scatter  of $\sigma$ = 1.68  m\,s$^{-1}$,
they must have the same filling factors to within 70\%.
Using the binned RV values in order to minimize the measurement error
of $\approx$ 2 m\,s$^{-1}$ then  the differences in amplitudes between
the RV curves of the four spot groups should be less 
than  $\sigma$ = 0.5  m\,s$^{-1}$. In this case
the spot filling factors for all spot groups must be the same to within
about 10\%. 

\item A symmetrical RV curve also implies that the spot evolution over the span
of the measurements for all four spot groups must be small.   The RV measurements
span more than 3.5 rotation periods and as noted by Pont et al.~(2010):
``no feature is reproduced unchanged after one rotation period, 
and the light curve becomes unrecognizable after merely 2-3 periods.'' 
Having four spot groups that evolve very little over 3.5 rotation periods
contradicts what we see in the light curve of CoRoT-7b.
\end{enumerate}

So, either we have a star with a superearth planet (similar to Kepler-10b - see below),
  or we have a star
with an extremely remarkable spot distribution. We thus conclude
that the 0.85-d RV modulation seen 
in Figure~\ref{corot7} and Figure~\ref{orbit} is due almost entirely to the reflex motion
induced by a planet.

We also see no evidence for any short term (i.e during the night)
systematic errors in the RV data (possible long term errors are removed by subtracting
the nightly mean). The error in each phase bin is just 
$\sigma_{RMS}$/$\sqrt{N}$ where $N$ are the number of points in each
bin used for averaging. If the HARPS RV errors are estimated well, then this
bin error should equal the mean RV error of the binned points divided
by $\sqrt{N}$. The last two columns in Table 2 compares these two values (signified by $\sigma_{obs}$, $\sigma_{cal}$).
The two values are consistent and the larger value was used as the error
bar in Figure~\ref{orbit}. 
We see no obvious evidence for short term systematic errors in the HARPS data.
This is also in agreement with RV planet search programs conducted with HARPS. 
There are several
dozen cases where the RV measurements of stars are in the same
signal-to-noise regime as CoRoT-7 and these show a residual scatter
in the range of 1.5--3.5 m\,s$^{-1}$ (Naef et al.~2010;
Lo Curto et al.~2010; Moutou et al.~2011), comparable to what we see in CoRoT-7.

Taking all the arguments into account, the most reasonable, and most logical
explanation for the 0.85-d RV variations of CoRoT-7  is that these are due 
entirely to the transiting planet, CoRoT-7b. Furthermore, the significance of
this detection is at the 15$\sigma$ level.

\section{The Bisector and FWHM Variations}

Queloz et al.~(2009) used a method similar to that of this paper to study the
possible variations of the cross correlation function bisector possibly induced by
the radial velocities variations due to the CoRoT-7b signal. For each of the
observing nights with two or three measurements per night, they corrected the radial
velocities from their average during the night over the two or three measurements.
They did the same correction on the bisector span by their nightly average. These
``differential'' bisector spans and radial velocities can show short-period
variations, on a time scale of about 5 hours which is the typical longest time
offset between data in night with multiple observations. This represents 25\%\ of
the CoRoT-7b orbit.

\begin{figure}
\plotone{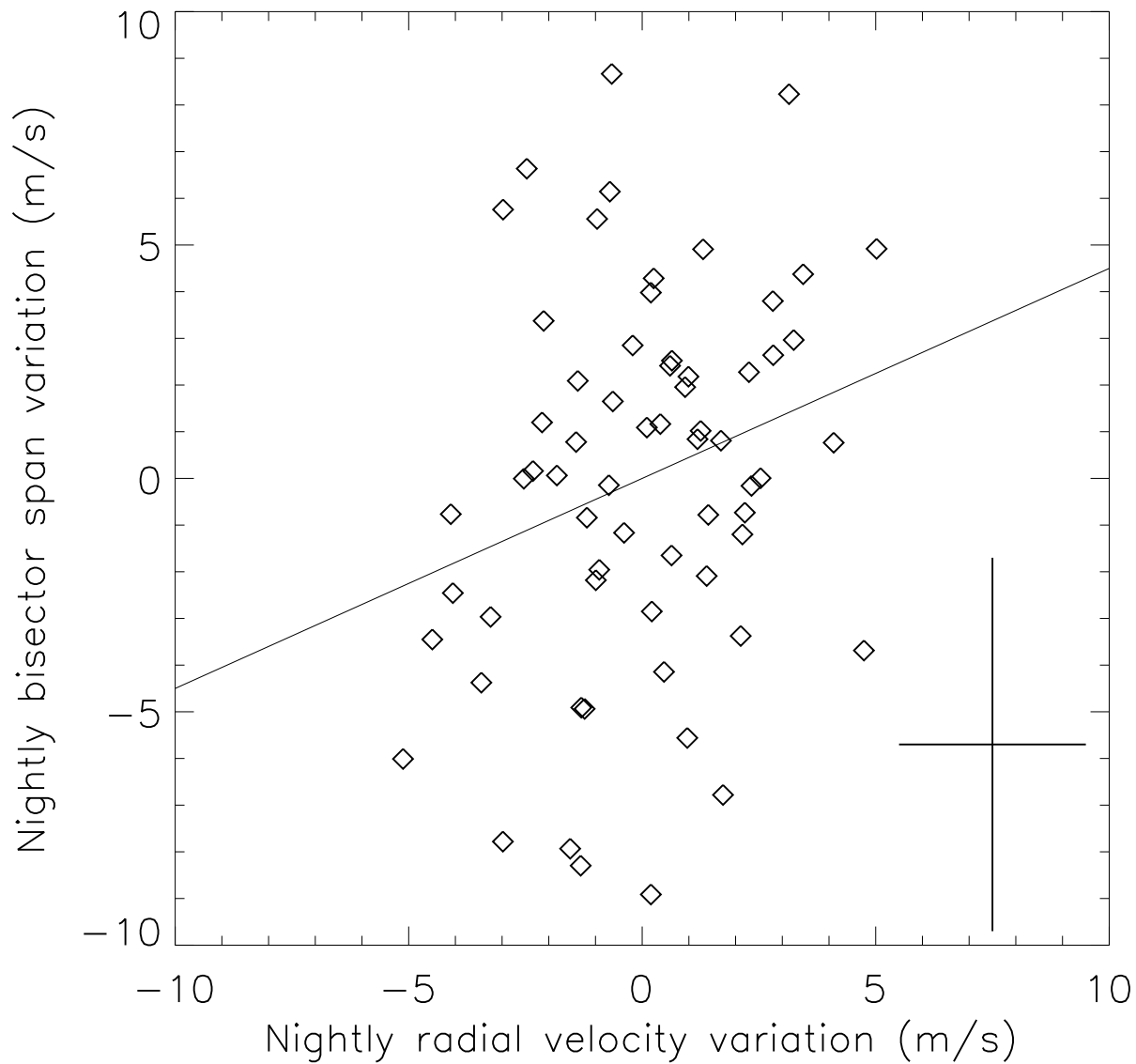}
\caption[]{
Bisector spans as a function of radial velocities (nightly variations). For each
observing night, both bisector spans and radial velocities are corrected for  the
average value of the night. The typical error bars on each point are plotted in the
bottom-right. There is a 2-$\sigma$ hint for the detection of a correlation between
the bisector spans and the radial velocities. The linear correlation has a slope of
$0.45\pm0.22$ that is shown on the plot.
\label{biscor}}
\end{figure}

The nightly variations of the bisector span as a function of the nightly variations
of the radial velocities are plotted in Figure~\ref{biscor}. It shows peak-to-peak variations on
the order of 20 m\,s$^{-1}$ for the bisector spans and 10   m\,s$^{-1}$
for the radial velocities. The
larger dispersion of the bisectors agrees with their error bars that are two times
larger than those for the radial velocities. The bisectors show a hint for
correlation with the radial velocities. With the hypothesis of a linear correlation,
the data show a positive slope of $0.45\pm0.22$ that is plotted in Figure~\ref{biscor} . The
Spearman's rank test indicates a 2$\sigma$ deviation from the null correlation
hypothesis. A bootstrap with 30\,000 pools mixing radial velocity and bisectors
shows a slope distribution centered on zero, with only 1.9\%\ probability to find a
slope of $0.45$ or larger. It also shows a 2.1\%\ probability to find by chance a
deviation from the null correlation hypothesis greater than 2$\sigma$ with the
Spearman's test. 

Thus, the data show a 2$\sigma$ detection of a positive correlation between the
differential bisector spans and radial velocities, on a night-per-night basis. Such
correlation could be the signature of blend scenario (e.g. Santos et al.~2002),
where a star in the background of the main target hosts a deep transit which seems
shallower as it is diluted by the constant, brighter flux of the main target.
However, blend simulations presented by Queloz et al.~(2009) have shown that the
slope the bisector versus the radial velocity should be steeper
than 2 in all cases, whereas $0.45\pm0.22$ is found here. Thus the possible
correlation between bisector spans and radial velocities apparently cannot be
explained with a blend scenario. In addition, such blend scenario would have to imply
that the background star hosts a deep transiting companion while the main star, by
chance, hosts at least one planet (CoRoT-7c) as detected in radial velocities. A
planetary system is more likely with the two companions orbiting the same stars.

\begin{figure}
\plotone{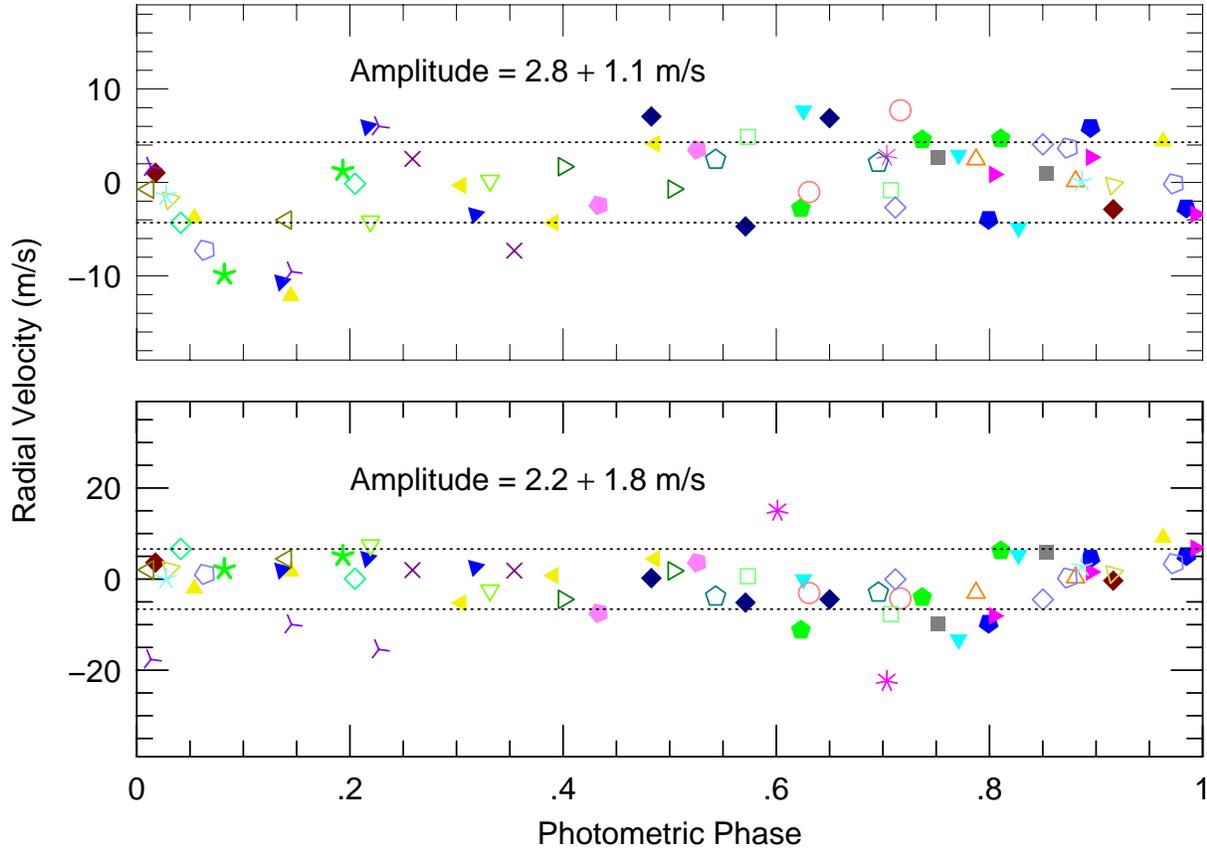}
\caption[]{
(Top) The residual bisector variations resulting from performing the same analysis as done
on the RV data and phased to the photometric period. The dashed horizontal lines represent
the $\pm 1 \sigma$ calculated for the data in the phase interval 0.15--1.0.
(Bottom) The residual FWHM resulting from performing the same analysis as done
on the RV data phased to the photometric period. 
\label{activity}}
\end{figure}

To investigate further any possible bisector variations, the exact
same analysis that
was applied to the RV  data was also applied to the bisector data. Namely,
the nightly data was treated as independent data sets and a sine wave was fit to the
data keeping the period fixed at 0.85 days and allowing the nightly means to float.
The top of the Figure~\ref{activity} shows the residual bisector variations after
subtracting the calculated nightly mean and phased to the CoRoT-7b period. There is no 
compelling evidence for sinusoidal variations. A sine fit to the
data yields an amplitude of 2.8 $\pm$ 1.1 m\,s$^{-1}$.
The only hint of variability is driven by  a cluster of only four points
at phase $\approx$ 0.1. Eliminating these points in a much lower amplitude
of 1.6 $\pm$ 0.9 m\,s$^{-1}$. 
Most
of the bisector span data between phases 0.2--1.0 have an rms scatter of 
$\pm$1$\sigma$ as shown by the dashed lines. 
For now any possible bisector correlation or variations
remain unexplained, but one should keep in mind that 
these variations with the present data set do not seem to be significant. 
Even though there are often large
nightly variations, these are all within the rms scatter of about
4.5 m\,s$^{-1}$. (Compared to the $\approx$ 10 m\,s$^{-1}$ rms scatter for the
bisector variations of Kepler-10b, see Figure 7 of Batalaha et al.~(2011).)
This also argues against any possible variations being due to a 
blend scenario.

We also applied this method to the FWHM measurements. This is shown in the bottom panel 
of Figure~\ref{activity} where we plot the FWHM residuals phased to the 0.85-d
period. The amplitude of any variations is 3.8 $\pm$ 3.2 m\,s$^{-1}$.
 There  is no
strong evidence for variability in the FWHM with a 0.85-d period.

\section{Discussion}

The planet mass can be derived from the $K$-amplitude via the mass function,
$f(m)$ which for circular orbits (the most likely case for CoRoT-7b) is:
\begin{equation}
f(m) = {{ (m_p sin i)^3 } \over {M_{star}^2}} = {{K^3P} \over {2\pi G}}
\end{equation}
where $P$ is the orbital period, $M_{star}$ the stellar mass, and $G$ the gravitational constant. 
Bruntt et al.~(2010) derived a stellar mass of
0.91 $\pm$ 0.03 $M_\odot$  and 
L\'eger, Rouan, Schneider et al.~(2009) derived an orbital inclination of $i$ = 80.1 $\pm$ 0.3 degrees.
Using the appropriate stellar and orbital parameters for CoRoT-7b and solving Eq. 1 for 
$m_p$ results in $m_p$ ($M_\oplus$) = (1.416 $\pm$ 0.031)$K$ (m\,s$^{-1}$). Our best fit
$K$-amplitude  of
5.25 $\pm$ 0.84 m\,s$^{-1}$ (linear model)
results in $m$ = 7.42 $\pm$ 1.21 M$_\oplus$. 

Given our derived planet mass we can now estimate the bulk density of the planet.
Bruntt et al.~(2010) give a planet radius of 1.58 $\pm$ 0.1 $R_\odot$. This 
results in a mean planet density of 
10.4 $\pm$  1.8 gm cm$^{-3}$.

The large density of CoRoT-7b has been reinforced by the discovery
of a second transiting superearth, Kepler-10b (Batalha et al.~2011).  
Kepler-10b has a slightly smaller mass,
$m$ = 4.56 $\pm$ 1.23 $M_\oplus$ 
and a smaller radius, $R$ = 1.416 $\pm$ 0.033. (For simplicity we 
 have taken the mean  of the
$\pm$ error values given in Batalha et al.~2011).  However, to within
the errors CoRoT-7b and Kepler-10b
have identical bulk densities:
$\rho_{Kepler-10b}$ = 8.8  $\pm$ 2.5 gm\,cm$^{-3}$ compared to 
$\rho_{CoRoT-7b}$ = 10.4  $\pm$ 1.8  gm\,cm$^{-3}$. 

\begin{figure}
\plotone{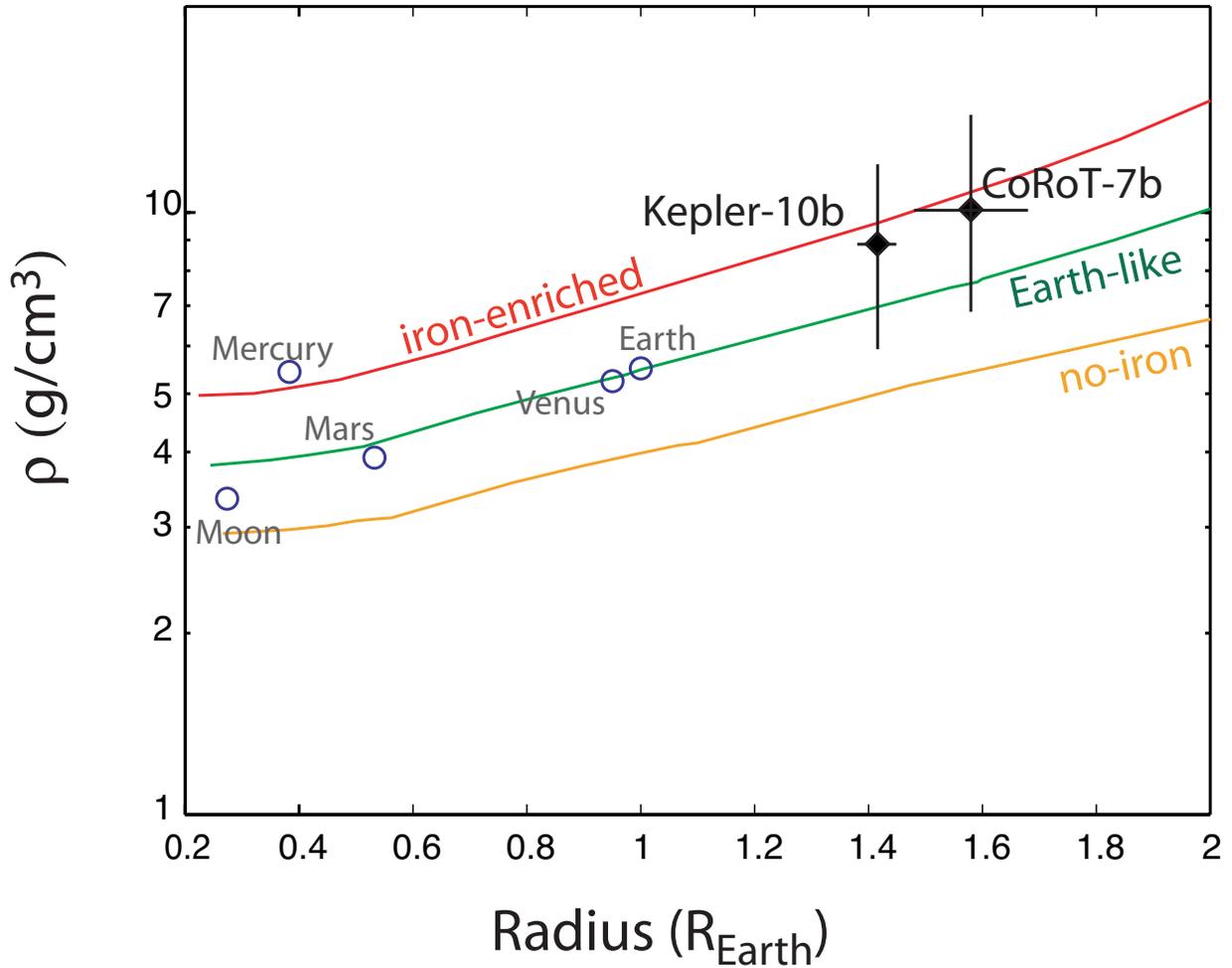}
\caption[]{
Density-Radius relationships for three types of rocky compositions. Earth-like: 67\% by mass of silicate 
mantle with 10\% iron by mol [(0.9MgO, 0.1Fe)+SiO$_2$] + 33\% by mass of iron core. No-iron: 100\% silicate mantle [MgO + SiO$_2$ mantle]. Iron-enriched: 37\% silicate mantle (with 0\% of iron by mol) + 63\% iron core. 
\label{rho}}
\end{figure}

 A detailed discussion on the possible internal structure of CoRoT-7b and
Kepler-10b is well beyond the scope of this paper. We can, however, 
compare the density and radius to terrestrial bodies in our own
solar system as well as to simple models. The inner terrestrial planets, excluding Mercury,
and the Moon show a tight
linear correlation between logarithm of the  density, $\rho$, and planet radius, $R_P$
(correlation coefficient, $r$ = 0.99976). The terrestrial planets as well as the moon
are shown in Figure~\ref{rho} along with
CoRoT-7b and Kepler-10b. Also shown are three models for a 
``super-Moon'', Earth-like, and a ``super-Mercury''. The Earth-like planet has a silicate
mantle that is 67\% by mass and an iron core that is 10\% by mass. A moon-like
planet has a 100\% silicate mantle (MgO and SiO$_2$) and no core. The Mercury-like
planet has an iron core that is 63\% of the mass and a silicate mantle 37\% of the
planet mass.

The nominal values of CoRoT-7b and Kepler-10b place these
above the Earth-like planets and close to Mercury that is iron enriched. 
However, within the errors
both of these exoplanets have a density and mass consistent with either an Earth-like
planet or a super-Mercury. Whether
CoRoT-7b and Kepler-10b have an internal structure similar to Mercury
or the Earth requires reducing  the error on the
 density. For CoRoT-7b equal contributions to the error
come from the planet mass (19\%) and radius (6\%, but contributing 
as $R^3$ in the density). Kepler-10b
benefits from a better measurement of the $R_p$ (due to the astroseismic
determination of the stellar parameters), but the planet mass is 
known to  within 26\%,  a bit worse than for
CoRoT-7b. A substantial improvement in the
mass of Kepler-10b can be realized by a reduction in the $K$-amplitude
error. The RV measurements for Kepler-10b show an rms scatter about the
orbit that is a factor of 2 larger than the internal errors. This is
most likely due to the RV jitter of activity 
(Batalha et al.~2011).
This activity jitter is estimated to be about
2 m\,s$^{-1}$, or 50\% of the planet $K$-amplitude. 
So, in order to get a better mass determination for
Kepler-10b one would have to correct for the activity signal, even for such
a quiet star.  Since Kepler-10b has a long rotational period ($P_{rot}$ = 50-100 d)
the stellar activity jitter could be reduced by
taking several measurements of Kepler-10 throughout the night when the star is well-placed
in the sky and employing the procedure we used
on CoRoT-7b. Unfortunately, the Kepler-10b RV measurements had only two
nights where multiple measurements were taken.

In light of the discovery of Kepler-10b, the properties of CoRoT-7b
may not be so extraordinary.
It is remarkable that so quickly after the start of the space missions CoRoT and Kepler
two transiting rocky planets with comparable (and possibly new) properties have 
been discovered. This
bodes well for the detection of more of such objects. 

 In the case of CoRoT-7b and Kepler-10b, the two host stars are also very similar 
with each having about the same radius, mass, and $T_{eff}$. The largest
 difference is in the level of activity  which is significantly lower in the case of Kepler-10. This is explained by Kepler-10 also being much older 
 than CoRoT-7, about 8 $\pm$ 0.26 Gyr (as derived by asteroseismology). CoRoT-7 has an
 estimated age of 1.2--2.3 Gyr 
 (L\'eger, Rouan, Schneider et al.~2009). The stars also differ somewhat in metalicity 
$[Fe/H]$ = +0.13 for CoRoT-7b (Bruntt et al.~2010) 
 and $[Fe/H]$ =  $-$0.1 for Kepler 10b (Batalha et al.~2011). Another similarity between the two systems is that of the very remarkable
 orbital period -- 0.85d -- for CoRoT-7b, which was unprecedented for such a low mass planet, is now
mirrored in Kepler-10b with a period of 0.84d.

 It should be noted here that if the mass of the exoplanet Kepler-9d (which is also orbiting a G-type star) is found to be close to the currently determined upper limit (7 M$_{\oplus}$, Holman et al~2010), this body also would belong to the same type of planet. This is not the case for the small planets recently discovered around the star Kepler-11 (Lissauer et al.~2011), which if their masses are confirmed belong to a type with very much lower densities 
($\approx$~0.8--2 gm\,cm$^{-3}$), so there appears to be a wide range of properties  of superearth  exo-planets. 
Figure~\ref{rho} hints that in terms of structure, CoRoT-7b and Kepler-10b
may be more like Mercury than the other terrestrial planets.
Clearly, more transiting superearths must be found - and with excellent
mass and radius determinations - before we know if CoRoT-7b and Kepler-10b
just represent a part of the ``continuum" of low mass planets, or whether
they are special even among short-period superearths.

Not surprisingly, the first confirmed superearths found by both CoRoT and Kepler are among
the brightest stars in their respective samples. Because they are relatively bright it was possible
to get the requisite precise RVs needed to confirm the nature of the transiting object.
There are undoubtedly more CoRoT-7b-like planets to be found in the CoRoT and Kepler samples. 
Unfortunately, these are most likely among the fainter stars of the sample for which  the
RV determination of the mass is challenging. Clearly, to make significant progress in the understanding
of the CoRoT-7b-type planets it is essential to find such objects around relatively bright stars
for which characterization studies are easier. 
Unfortunately, the requisite precise 
photometry can only be conducted from space. This only emphasizes the need for such proposed
space missions as PLAnetary Transits and Oscillations (PLATO, see Catala~(2009)) and 
the Transiting Exolanet Survey Satellite (TESS, see Ricker et al.~2009)
 whose goals are to find transiting terrestrial planets
among the brightest stars. In the case of PLATO, stellar parameters will be determined
precisely using astroseismology which translates into more accurate planetary parameters.
We will thus be able to determine if the ``CoRoT-7b-like" planets are indeed more like
Mercury than the Earth in terms of their internal structure.

\vskip 0.1in

\centerline{Acknowledgments}

The German CoRoT Team (Th\"uringer Landessternwarte and University of Cologne)
acknowledges the support of grants 50OW0204, 50OW603, and 50QM1004 
from the Deutsches Zentrum f\"ur Luft- und Raumfahrt e.V. (DLR). TM and GN acknowledge the
support of the Israel Science Foundation (grant No. 655/07).

\clearpage

\begin{deluxetable}{ccccr}
\tablecaption{Journal of Observations}
\tablewidth{0pt}
\tablehead{
\colhead{Start}         & \colhead{$N_{obs}$ } & \colhead{$\Delta T$}  & \colhead{$\Delta\phi$}  &   \colhead{$\Delta V_{C7c,d}$}  \\
  &     & \colhead{(hours)}  &     &   \colhead{(m\,s$^{-1}$}) }
\startdata
2454799.7770 & 2 &  2.10 & 0.10 &  $-$0.33   \\
2454800.9461 & 2 &  2.09 & 0.10 &  0.05 \\
2454801.7530 & 2  & 2.08 & 0.10 &  0.76 \\
2454802.7484 & 2  & 2.27 & 0.11 &  $-$0.45 \\
2454803.7526 & 2 & 1.95 & 0.09 & $-$0.95 \\
2454804.7554 & 2 & 1.75 & 0.09 &  $-$0.07\\
2454805.7775 & 2 & 1.75 & 0.09 &  $-0.35$  \\
2454806.7647 & 2 & 1.91 & 0.09 &  $-$0.42\\
2454807.7281 & 2 & 2.36 & 0.12 &  $-$0.18 \\
2454847.5968 & 3 & 3.84 & 0.19 &  $-$1.63 \\
2454848.6008 & 3 & 3.80 & 0.19 &  $-$0.78\\
2454849.5940 & 3 & 3.72 & 0.18 &  0.76 \\
2454850.5951 & 3 & 3.72 & 0.18 &  $-$0.52 \\
2454851.5927 & 3 & 3.72 & 0.18 &  1.19\\
2454852.7986 & 3 & 3.42  & 0.17 &   $-$0.17 \\
2454853.5740 & 3 & 4.13 & 0.20 &  2.00 \\
2454854.5802 & 3 & 3.88 & 0.19 &  $-$0.27\\
2454865.5978 & 2 & 2.83 & 0.14 &   0.01 \\
2454867.5604 & 2 & 2.30 & 0.11 &  0.17 \\
2454868.5918 & 2 & 2.30 & 0.11 &  0.32 \\
2454869.6002 & 2 & 2.11 & 0.10 &  $-0.63$ \\
2454870.6013 & 2 & 2.75 & 0.13 &  $-0.07$\\
2454872.5647 & 3 & 3.91 & 0.19 & $-$1.43  \\
2454873.5375 & 3 &  4.38 & 0.21 & 0.96 \\
2454879.6002 & 2 & 3.35 & 0.16 &  1.33 \\
2454882.5256 & 2 & 3.11 & 0.15 &  1.16 \\
2454884.5265 & 2 & 2.87 & 0.12 &  $-$1.43\\
\enddata
\end{deluxetable}

\begin{deluxetable}{crccc}
\tablecaption{Binned RV values}
\tablewidth{0pt}
\tablehead{
\colhead{$\phi$}          & \colhead{RV}              & \colhead{$N_{obs}$} & 
\colhead{$\sigma_{obs}$}  & \colhead{$\sigma_{cal}$}  \\
                          & \colhead{(m\,s$^{-1}$)}   &  
                          & \colhead{(m\,s$^{-1}$)}   & \colhead{(m\,s$^{-1}$)}
} 
\startdata
0.00  &   0.01   &  7   & 0.70   & 0.68 \\
0.05  & $-$2.03  &  6   & 0.93   & 0.80 \\
0.14  & $-$3.65  &  4   & 1.06   & 0.93 \\
0.22  & $-$4.89  &  6   & 0.37   & 0.80 \\
0.36  & $-$3.53  &  7   & 0.46   & 0.75 \\
0.56  &    1.22  &  11 & 0.52   & 0.53 \\
0.71  &    5.61  &  8   & 0.51   & 0.74 \\
0.80  &    5.72  &  6   & 0.50   & 0.65 \\
0.88  &    2.88  & 9    & 0.76   & 0.60 \\
\enddata
\end{deluxetable}

\begin{deluxetable}{clcll}
\tablecaption{Mass Determinations for CoRoT-7b}
\tablewidth{0pt}
\tablehead{
\colhead{$N$}   & \colhead{K-amp (m\,s$^{-1}$)}  & \colhead{Mass (M$_\odot$)} & 
\colhead{Method} &
\colhead{Reference}    } 
\tablehead{
\colhead{$N$}   & \colhead{K-amplitude}  & \colhead{Mass} & 
\colhead{Method} & \colhead{Reference}    \\
     & \colhead{(m\,s$^{-1}$)}  & \colhead{(M$_\oplus$)} &  &
} 
\startdata
1 & 5.70  $\pm$ 0.80 & 8.0  $\pm$ 1.20  & High Pass Filtering & Ferraz-Melo (2011) \\
2 & 5.15 $\pm$ 0.95 & 7.29 $\pm$ 1.35 & Model independent  & This work \\
3 & 5.27 $\pm$ 0.81 & 7.46 $\pm$ 1.16 & Linear model  & This work \\
4 & 5.04 $\pm$ 1.09 & 7.12 $\pm$ 1.43 & Offset fitting & Hatzes et al. (2010) \\
5 & 4.16 $\pm$ 1.00 & 5.88 $\pm$ 1.49 & Pre-whitening & Queloz et al. (2009) \\
6 & 4.00 $\pm$ 1.60 & 5.65 $\pm$ 1.61 & Harmonic Filtering & Boisse et al. (2011) \\
7 & 3.80 $\pm$ 0.80 & 5.37 $\pm$ 0.82 & Harmonic Filtering  w/correction & Queloz et al. (2009) \\
8 & 3.50  $\pm$ 0.60 & 4.96  $\pm$ 0.86  &  Adopted & Queloz et al. (2009) \\
9 & 3.33  $\pm$ 0.80 & 4.70  $\pm$ 1.40  &  Pre-whitening w/correction & Queloz et al. (2009) \\
10 & 1.90 $\pm$ 0.40 & 2.65 $\pm$ 0.56 & Harmonic Filtering & Queloz et al. (2009) \\
11 & 1.60 $\pm$ 1.30 & 2.26 $\pm$ 1.83 & Activity Modeling & Pont et al. (2010) \\
\enddata
\end{deluxetable}

\end{document}